\documentclass[%
reprint,
 amsmath,amssymb,
 aps,
prl,
superscriptaddress,
]{revtex4-2}

\usepackage{graphicx}% Include figure files
\usepackage{dcolumn}% Align table columns on decimal point
\usepackage{array}   % for \newcolumntype macro
\newcolumntype{L}{>{$}c<{$}} % math-mode version of "l" column type
\usepackage{bm}% bold math
\usepackage[justification=raggedright]{caption}
\usepackage{subcaption}
\usepackage{physics}
\usepackage{algorithm}
\usepackage{algorithmic}
\usepackage{comment}
\usepackage{xcolor} % use color in text
\usepackage{hyperref}% add hypertext capabilities
\hypersetup{colorlinks = true,linkcolor = blue, breaklinks = true}

\DeclareMathOperator{\erfc}{erfc}
\newcommand{\fig}{Fig.~}
\newcommand{\eq}{Eq.~}

\newcommand{\qpar}{q_{||}}
\newcommand{\sbar}{\hat{s}}
\newcommand{\euler}{\text{e}}
\newcommand{\lqN}{\lambda_{near}}
\newcommand{\lqF}{\lambda_{far}}
\newcommand{\lqS}{\lambda_{single}}
\begin{document}

% \preprint{APS/123-QED}

\title{Experimental observations of bifurcated power decay lengths in the near Scrape-Off Layer of tokamak plasmas}% Force line breaks with \\
%\thanks{A footnote to the article title}%

\author{X. Zhang}
\email{laura.zhang@tokamakenergy.com}
\author{C. Marsden}
\author{M. Moscheni}
\author{E. Maartensson}
\author{A. Rengle}
\author{M. Robinson}
\affiliation{Tokamak Energy Ltd, Abingdon, United Kingdom}

\author{T. O'Gorman}
\affiliation{Tokamak Energy Ltd, Abingdon, United Kingdom}

\author{H. F. Lowe}
\author{E. Vekshina}
\author{S. Janhunen}
\thanks{author's current address: Los Alamos National Laboratory, New Mexico 87545, United States}
\affiliation{Tokamak Energy Ltd, Abingdon, United Kingdom}

\author{A. Scarabosio}
\affiliation{Tokamak Energy Ltd, Abingdon, United Kingdom}
\affiliation{LINKS Foundation, 10138 Turin, Italy}

\author{P. F. Buxton}
\author{M. Sertoli}
\author{M. Romanelli}
\author{S. McNamara}
% \author{the ST40 Team}
\affiliation{Tokamak Energy Ltd, Abingdon, United Kingdom}

\author{T. K. Gray}
\affiliation{Oak Ridge National Laboratory, TN, United States of America}

\author{N. A. Lopez}
\affiliation{University of Oxford, Oxford, United Kingdom}

\collaboration{the ST40 Team}
\noaffiliation

% \date{\today}% It is always \today, today,

\begin{abstract}
The scrape-off layer parallel heat flux decay lengths measured at ST40, a high field, low aspect ratio spherical tokamak, have been observed to bifurcate into two groups. The wide group follows established H-mode scalings while the narrow group falls up to 10 times below these scalings, being comparable to the ion Larmor radius rather than the ion poloidal Larmor radius. The onset of the narrow scrape-off layer width is observed to be associated with suppressed magnetic fluctuations, suggesting reduced electromagnetic turbulence levels in the SOL.
\end{abstract}

%The wide group matches closely with the scale of ion poloidal Larmor radius and follows existing H-mode scalings, while the narrow group falls up to 10 times below scalings, on the scale of ion total Larmor radius.

\maketitle

\paragraph{Introduction.}
Nuclear fusion promises a clean and abundant source of energy. A nuclear fusion reactor is expected to require tens of megawatts of heating power that brings the confined plasma to nuclear fusion temperatures, which in turn could produce up to gigawatts of fusion power. This intense amount of energy is stored in charged particles, and transported by these charged particles from the center of the fusion plasma to the edge, the so-called scrape-off layer (SOL), where the magnetic field lines are open and intersect the material surfaces. Because of the narrow width of this layer, the heat flux density that arrives at the material wall can be well above 20~MW/m$^2$, which, even with advanced mitigation methods \cite{antar2010convective, lunt2023compact, nelson2023robust}, is extremely challenging for the plasma facing components to handle \cite{fasoli2023essay}. This challenge can be exacerbated by absorption of auxiliary heating power in the SOL itself \cite{perkins2012high} or the occurrence of edge localized modes (ELMs) \cite{nelson2023robust, suttrop2011first}. The plasma heat flux handling challenge has therefore been one of the main focus areas of the fusion plasma physics community \cite{fasoli2023essay}. 

Being able to predict the width of the heat flux channel and the shape of its profile are crucial for designing divertors for the next generation reactors that ensures survivability. Current methods of predicting the width of the near SOL heat flux channel - henceforth referred to as ``SOL width" - are based mainly on single and multi-machine scaling laws constructed with statistical regression of measured SOL widths against combinations of input parameters. Notable examples include the ITPA multi-machine scalings by Eich \textit{et al} \cite{eich2013scaling} and the ASDEX edge pressure scaling by Silvagni \textit{et al} \cite{silvagni2020scrape}, which was then improved upon and confirmed at Alcator C-MOD and JET \cite{ballinger2022dependence, faitsch2020correlation}.

The Heuristic Drift (HD) model by Goldston \textit{et al} \cite{goldston2011heuristic, eich2011inter, goldston2015theoretical}, which is similar to regression \#14 in \cite{eich2013scaling}, provides a physical interpretation of the near SOL width: the width of the channel is set by the neoclassical excursion of ions into the SOL, with electron anomalous transport ensuring ambipolarity \cite{goldston2011heuristic}. This interpretation is consistent with results from gyrokinetic simulations for present-day machines by XGC \cite{chang2017gyrokinetic}. Moreover, XGC simulations predicted that if the electron anomalous transport is enhanced, the electron heat flux will dominate over that of neoclassical ions, ultimately broadening the SOL width. Such turbulence-induced broadening was predicted to become measurable at higher currents for STs and next generation devices \cite{chang2017gyrokinetic}. This broadening effect was subsequently observed at ASDEX and DIII-D \cite{eich2020turbulence, ernst2024broadening, faitsch2021broadening}, giving confidence to the wider SOL widths predicted for next-generation devices including ITER.

However, recent publications from TCV \cite{maurizio2021h} and COMPASS \cite{hecko2023experimental} for values of poloidal magnetic field at the outer mid-plane $B_{p}\approx 0.2$~T, similar to NSTX and MAST \cite{eich2013scaling}, reported SOL widths up to 5 times below the Eich regression \#14 in \cite{eich2013scaling}. 
% For both machines, it was reported that Eich regression \#9, which included explicit dependencies on toroidal field $B_t$, $q_{95}$, and input power and performed on conventional tokamaks only, instead gave the closest match to data. 
The reason behind this deviation from the ITPA database remains unexplained, especially the weakened predictive power of $B_p$. Here we present a set of measurements from the ST40 spherical tokamak \cite{mcnamara2023achievement} which reproduces both the ITPA-like and the COMPASS-like SOL widths, and demonstrates that there are actually two bifurcated branches of near SOL transport, with the resulting SOL widths coinciding with the ion poloidal and total Larmor radii, respectively.

\paragraph{The ST40 SOL width database.}
ST40 is a compact high-field spherical tokamak (ST) constructed and operated by Tokamak Energy Ltd, located near Oxford, Oxfordshire, UK, with major radius $R = 0.4 - 0.5$~m, minor radius $a = 0.2 - 0.25$~m, and on-axis magnetic field up to 2~T. With aspect ratio similar to MAST and NSTX, but magnetic fields and safety factors closer to the conventional tokamaks TCV and COMPASS, ST40 is uniquely positioned to disentangle the effect of device size and field strength versus aspect ratio on the SOL width. In the 2023 campaign, ST40 achieved the diverted configuration, along with significant upgrades in diagnostics both in the core and in the upper outer divertor \cite{mcnamara2024overview}, enabling such a study. 

\begin{table}[]
    \centering
    \begin{tabular}{L | L | L | L | L}
        \hline
        B_t \text{~(T)} &  I_p \text{~(kA)} & P_{tot} \text{~(MW)} & \Bar{p}_e \text{~(kPa)} & q_{95}\\
        \hline
         0.7 - 1.5 & 300 - 550 & 0.5 - 2 & 10 - 19 & 4 - 15\\
         \hline 
    \end{tabular}
    \caption{Selected parameters of the ST40 scrape-off width dataset, with $B_t$ the on-axis toroidal field, $I_p$ the plasma current, $P_{tot}$ the total (beam + Ohmic) heating power, $\Bar{p}_e$ the line-averaged electron pressure, and $q_{95}$ the edge safety factor.}
    \label{tab:parameters}
\end{table}

%Selected engineering parameters of the ST40 scrape-off width dataset, including the on-axis toroidal field $B_t$, plasma current $I_p$, total heating power including injected beam power and Ohmic heating $P_{tot}$, line averaged electron pressure $\Bar{p}_e$, and edge safety factor $q_{95}$.

A dataset has been assembled for all ST40 diverted discharges with divertor infrared (IR) camera measurements of the divertor surface temperature. The camera resolution is $640 \times 512$ pixels, providing 0.2~mm spatial resolution on the divertor, with images acquired at 800~Hz \cite{PSI_MR, PSI_MM}. After selecting for data quality, the final database consists of 351 heat flux profiles across 33 plasma discharges, and is augmented with the electron temperature and density measurements from Thomson Scattering (TS), as well as magnetic fluctuation data collected by fast poloidal field probes located behind the upper divertor (measuring the $B_p$ component perpendicular to the flux surfaces). The discharges span a wide range of engineering parameters, as shown in Tab.~\ref{tab:parameters}.
Only time steps with plasma current within 10\% of the flat top values are included in the data base. The shape of the plasma is similar across all shots, with inverse aspect ratio $\epsilon \sim 0.6$ and elongation $\kappa \sim 1.8$. All discharges were nominally double-null diverted, with distance between the two separatrices $0> dr_{\text{sep}}>-3$~mm at the OMP according to EFIT reconstruction, biased towards the lower target (unfavorable for H-mode access). All beam heated shots showed strong density pedestals with no improvement in confinement observed. No large edge localized modes (type-I) were observed in the pulses in the database. 

The temperatures measured by the IR camera are captured as a time series of 2D images, and sets of heat flux profiles in the cross-field (radial) direction that are free of surface hot spots and away from tile edges are selected for heat flux calculations. The surface heat flux is calculated by solving the time-dependent homogeneous heat equation constrained by the measured surface temperature data in two dimensions, including one dimension along the tile surface and another into the tiles \cite{PSI_MR, PSI_MM}. The results from such 2D analysis have been compared against 3D calculations, and shown no significant discrepancy in regions of interest\cite{PSI_MR, PSI_MM}.

\begin{figure}
    \centering
    \includegraphics[width = 0.49\linewidth]{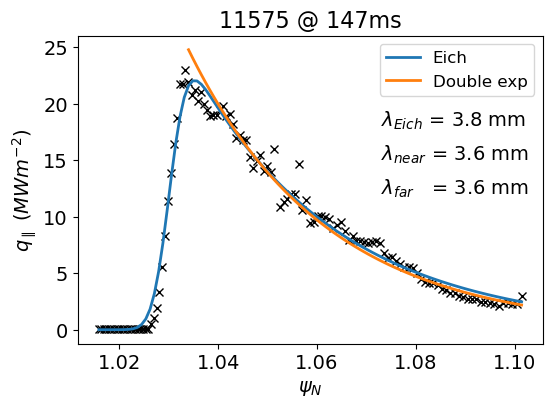}
    \includegraphics[width = 0.49\linewidth]{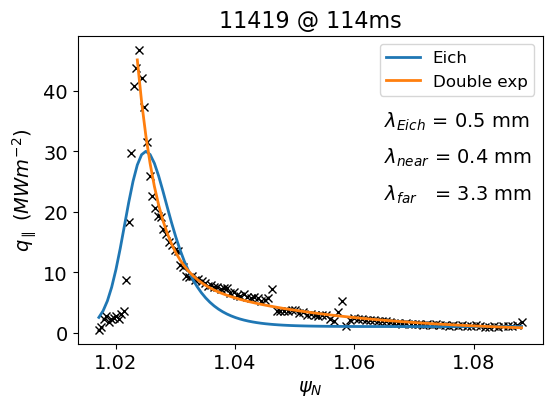}\\
    \includegraphics[width = 0.49\linewidth, trim = {4mm 6mm 4mm 6mm}, clip]{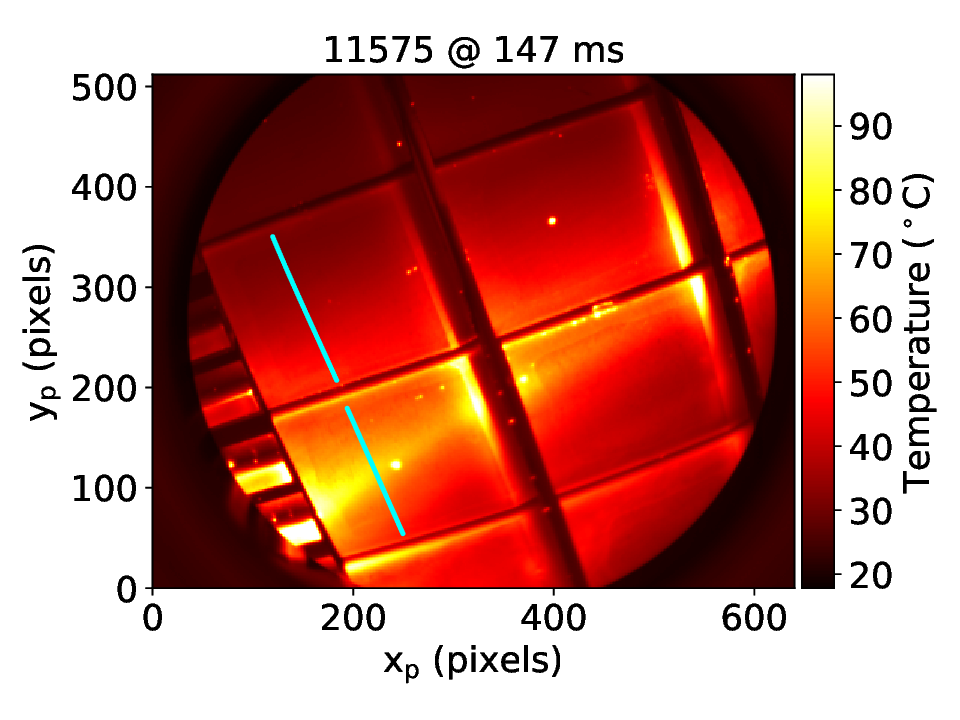}
    \includegraphics[width = 0.49\linewidth, trim = {4mm 6mm 4mm 6mm}, clip]{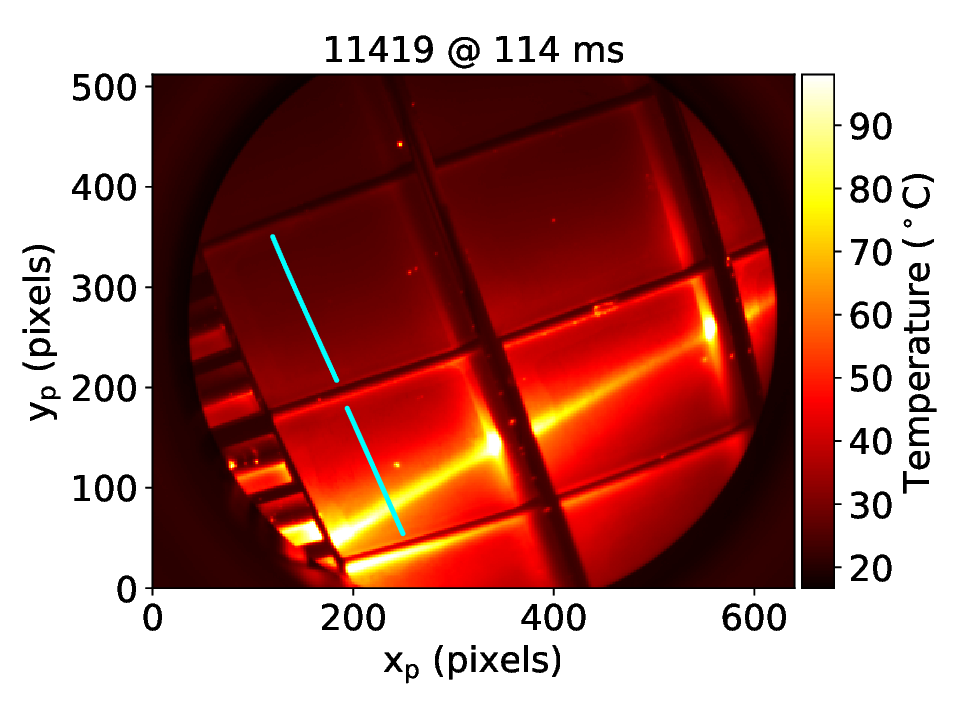}

    \caption{(Top) Examples of (left) a single-exponential profile and (right) a double-exponential profile for the parallel heat flux in two different discharges, along with the results of profile fits from both \eq\ref{eq:Eich} and \eq\ref{eq:Brunner}. The multi-exp function of \eq\ref{eq:Brunner} can robustly fit to both types of profiles. (Bottom) Raw IR camera image, with the profiles above corresponding to the indicated cyan line.}
    \label{fig:twofits}
\end{figure}

The calculated surface heat flux profiles are mapped to the outer mid-plane (OMP) in three dimensions using the Computer Aided Design (CAD) model of the ST40 inner vacuum chamber and EFIT reconstructed magnetic equilibrium to obtain the parallel heat flux profile $\qpar(\rho)$, where $\rho$ is the poloidal flux coordinate \cite{PSI_CM}. The remapped profiles are then fitted to two non-linear functional forms: a single exponential function in the common flux region (CFR) convolved with a Gaussian function to account for the heat flux spreading \cite{eich2013scaling},
\begin{multline}
     \qpar(\sbar) = q_{BG} + \frac{q_0}{2}\exp\qty[\qty(\frac{S}{2\lambda_{Eich}})^2 - \frac{\sbar}{\lambda_{Eich}}]\\ \times\erfc\qty(\frac{S}{2\lambda_{Eich}} - \frac{\sbar}{S})
     ,
     \label{eq:Eich}
\end{multline}
where $S$~(mm) is the spreading factor,  $\lambda_{Eich}$~(mm) is the heat flux width, and $q_{BG}$ is the background heat flux; and a double-exponential function for the CFR \cite{brunner2018high},
\begin{equation}
     \qpar(\sbar) =
     (q_0 - q_{\text{cf}}) \euler^{-\sbar / \lqN} + q_{\text{cf}} e^{-\sbar / \lqF}
     ,% \hspace \sbar \geq 0
    \label{eq:Brunner}
\end{equation}
where $q_{cf}$~(MW/m$^2$) is the peak heat flux of the `far' SOL exponential, and $\lqN$~(mm) and $\lqF$~(mm) are the `near' and `far' heat flux width respectively. In both functional forms $q_0$ is the upstream peak heat flux, and $\sbar$~(mm) denotes the radial coordinate along the outer mid-plane with $\sbar = 0$ at the separatrix. The analysis is repeated across the tile surface at different toroidal locations, and the resulting heat flux widths are averaged, with the standard deviation measuring the spread. Only results with standard deviation less than 20\% of the mean are kept as good quality measurements.

%The decay lengths in the PFR is not studied in this work.

Examples of both fits are shown in \fig\ref{fig:twofits}. It can be seen that while the Eich fit \ref{eq:Eich} gives satisfactory results when the CFR consists of a single exponential decay, there are cases when two exponentials are clearly seen in the CFR that can only be successfully fitted by Eq.~\ref{eq:Brunner}. When the Eich fit is successful, however, the heat flux profile can be well-described by a single exponential function, and indeed, the $\lqN$ and $\lqF$ found by the multi-exp fits are equal to each other. For this reason, the decay lengths from the multi-exp fit are used for the rest of this paper, with the two decay lengths in the CFR henceforth referred to as `double, near' (or $\lqN$) for the near SOL , and `double, far' (or $\lqF$) for the far SOL. When the two widths are equal, they will be referred to as $\lambda_{single}$ for simplicity. Note that $\lambda_{single} \approx \lambda_{Eich}$.

%$\lambda_{single} \equiv \lqN \equiv \lqF \approx \lambda_{Eich}$ is used.

\begin{figure}
     \centering
     \begin{subfigure}{0.49\linewidth}
         \centering
         \includegraphics[width = \linewidth]{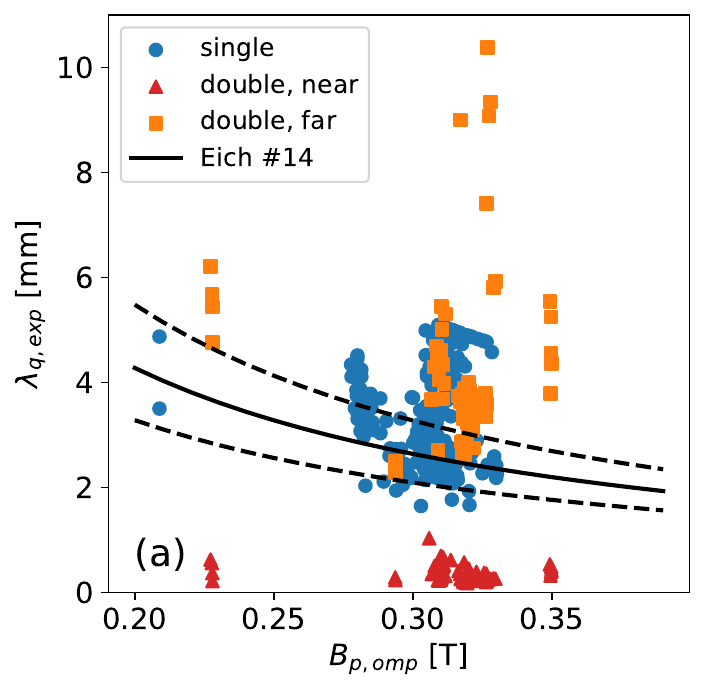}
         % \caption{Eich \#14}
         \label{fig:Eich14}
     \end{subfigure}%
     \begin{subfigure}{0.49\linewidth}
         \centering
         \includegraphics[width = \linewidth]{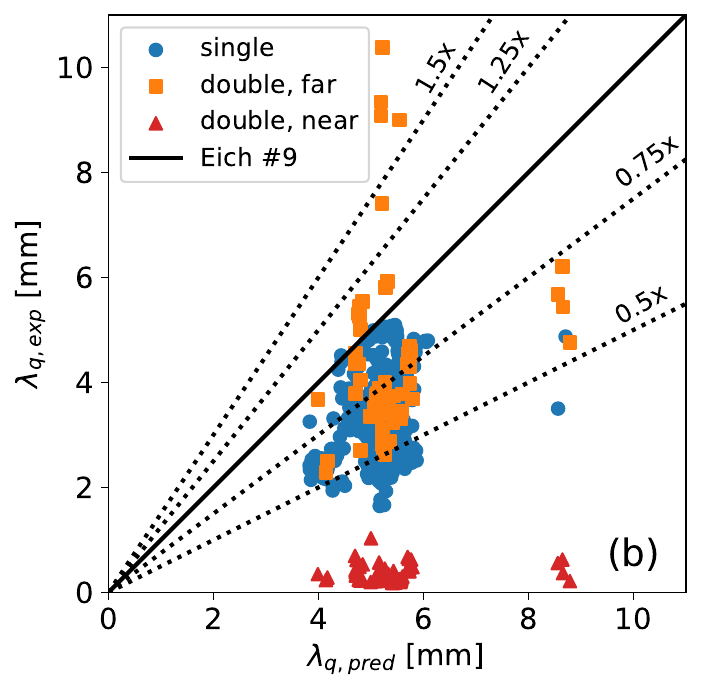}
         % \caption{Eich \#9}
         \label{fig:Eich9}
     \end{subfigure}
     \begin{subfigure}{0.49\linewidth}
         \centering
         \includegraphics[width =\linewidth]{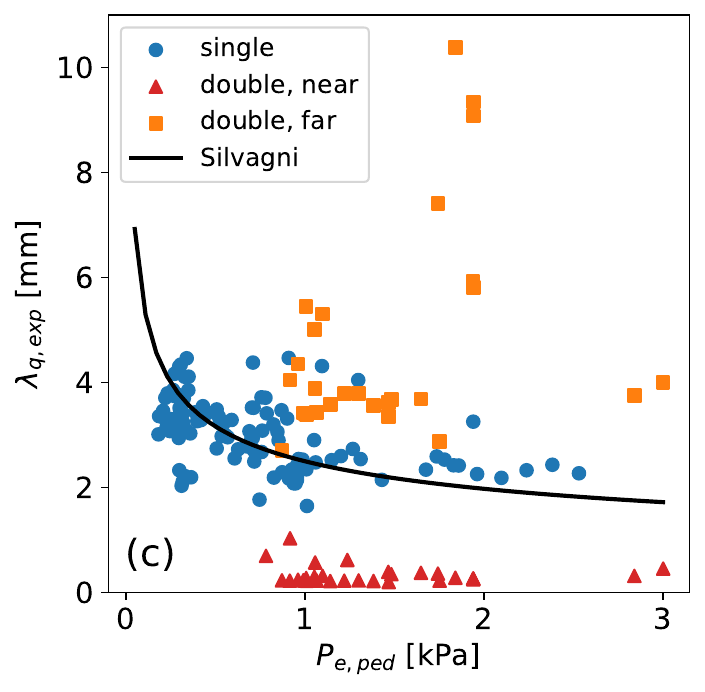}
         % \caption{Silvagni scaling}
         \label{fig:Silvagni}
     \end{subfigure}%
     \begin{subfigure}{0.49\linewidth}
         \centering
         \includegraphics[width =\linewidth]{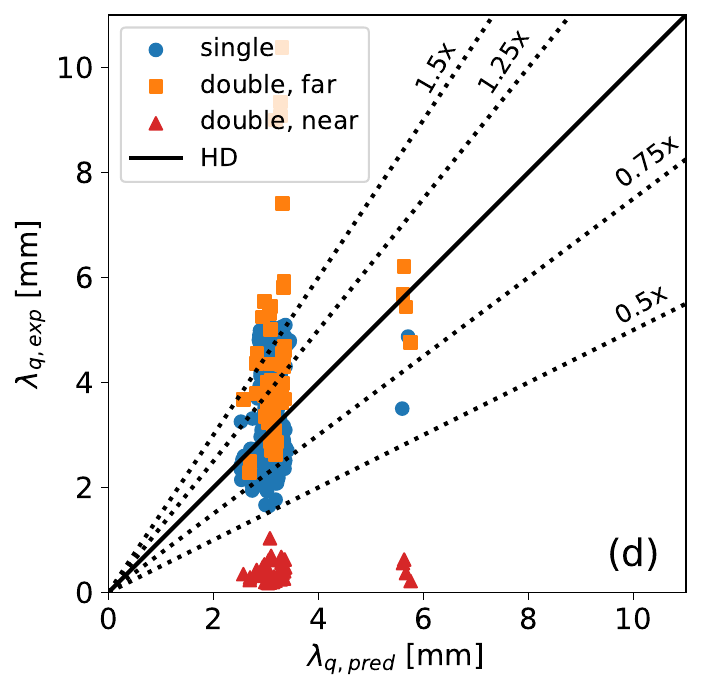}
         % \caption{HD model}
         \label{fig:HD}
     \end{subfigure}
    \caption{Comparison of experimentally measured SOL power decay length with existing scalings. Measured $\lqS$ and $\lqF$ fall into a single wide branch, with the onset of narrow branch $\lqN$ occurring at all poloidal field values, and moderate to high pedestal pressures.}
    \label{fig:scalings}
\end{figure}

\paragraph{Bifurcated SOL widths.}
Figure~\ref{fig:scalings} compares the measured ST40 SOL widths in single-exponential profiles (having only $\lqS$) and in double-exponential profiles (having both $\lqN$ and $\lqF$) against select scalings (a)-(c)\cite{eich2013scaling, silvagni2020scrape}, as well as against the heuristic drift model (d) \cite{goldston2011heuristic, goldston2015theoretical}. Electron pressure at the top of the density pedestal is used instead of that at normalized poloidal flux $\Psi_N = 0.95$ for the Silvagni scaling, similar to \cite{faitsch2020correlation}. Two distinct branches of SOL widths are clearly seen in \fig\ref{fig:scalings}. The wide branch consists of $\lqS$ and $\lqF$, the majority of which follows existing H-mode scalings to $\pm 50\%$. Qualitatively, the dependency on $B_p$ agrees with scaling predictions, Fig.~\ref{fig:scalings}(a) and (d), but the dependency on $p_{e, ped}$ is weaker than scaling, especially at higher pressures. The narrow branch consists simply of $\lqN$ and is not predicted by any of the established scalings.

In previous work, it is argued that $\lqN$ would follow the same physics that determines $\lqS$ (neoclassical ion excursion into the SOL), and $\lqF$ would be set by intermittent, turbulent transport \cite{goldston2015theoretical, brunner2018high}; this principle has been corroborated by various tokamaks operated in both limited and diverted configurations \cite{goldston2015theoretical, brunner2018high, ballinger2022dependence}, successfully comparing $\lqN$ and $\lqS$ in both magnitude and dependencies. The observation here that $\lqF$ in ST40 coincides with $\lqS$, rather than $\lqN$, stands in contrast with these previous findings. This suggests that in ST40, $\lqF$ is dominated by the neoclassical ion drifts, while $\lqN$ is governed by a manifestly different transport mechanism that remains to be fully understood. %This leads to the important conclusion that the near SOL transport process is indeed bifurcated into two branches. 

\paragraph{Narrow branch and reduced fluctuations.}
The width of the narrow branch clusters around $\lqN \sim 0.3$~mm, falling up to 10 times below existing scalings. The narrow branch is observed at all fields and currents, but only in beam-heated plasmas. Continuous observation of double-exponential heat flux profiles has been seen in time windows up to 15~ms (which is comparable to plasma energy confinement time), but only in full field, high current ($I_p > 450$~kA, $B_p > 0.3$~T) discharges. During such continuous windows, the plasma pedestal pressure is also observed to rise continuously, followed by an immediate crash when $\lqN$ returns to the same magnitude as $\lqS$. When the narrow peak persists for longer than 20~ms, the IR camera saturates (at 100 degree Celsius) due to the intense heating of the targets and rapid rise in tile temperature. It is therefore not possible to conclude whether the double-exponential profiles persist only up to a few energy confinement time, or are indeed a steady-state phenomena. Improvement in global confinement that would signal L-H transitions was not consistently observed during the bifurcation windows.

\begin{figure}
    \centering
    \includegraphics[width = 0.9\linewidth]{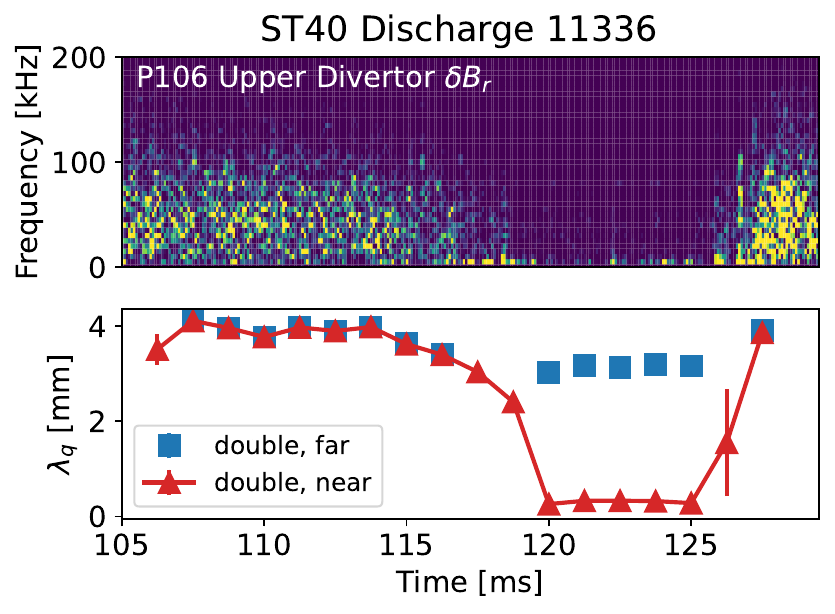}
    \caption{An example of a transient onset of narrow near SOL width. (top) Spectrogram of $dB/dt$ signals from poloidal magnetic pickup coil, located at the upper divertor measuring $B_r$. (bottom) Measured $\lqN$ and $\lqF$. The onset of the narrow branch coincides with the period of suppressed magnetic fluctuations, between 118 and 125~ms.}
    \label{fig:11336}
\end{figure}

An example of a transient onset of the narrow branch is shown in \fig\ref{fig:11336}, where the transition between the wide branch and narrow branch was observed. The spectrograms from the magnetic pick-up loop located behind the upper outer divertor target, oriented to measure $B_r$, the component of poloidal magnetic field perpendicular to the flux surfaces, is shown alongside the SOL width measurements during the same time window. A reduction in fluctuation intensity in the low frequency range of $<100$~kHz can be seen in the magnetic signal, which coincides with the bifurcation of $\lqS$ into $\lqN$ and $\lqF$. Pedestal temperature and density measurements from TS show an increase in pedestal pressure through 107, 117, and 127ms, followed by an immediate crash at 137~ms. Both the sudden increase of SOL width after 127~ms and the reduction in pedestal pressure immediately after indicate a strong increase in edge transport. %The change in $B_r$ fluctuation levels occurs on the microsecond time scale, with the $\lqN$ responding in $\sim1$~ms, consistent with the timescales of plasma turbulence and SOL transport respectively. 

\begin{figure}
    \centering
    \includegraphics[trim={0.28cm 0 0 0},clip, width = 0.8\linewidth]{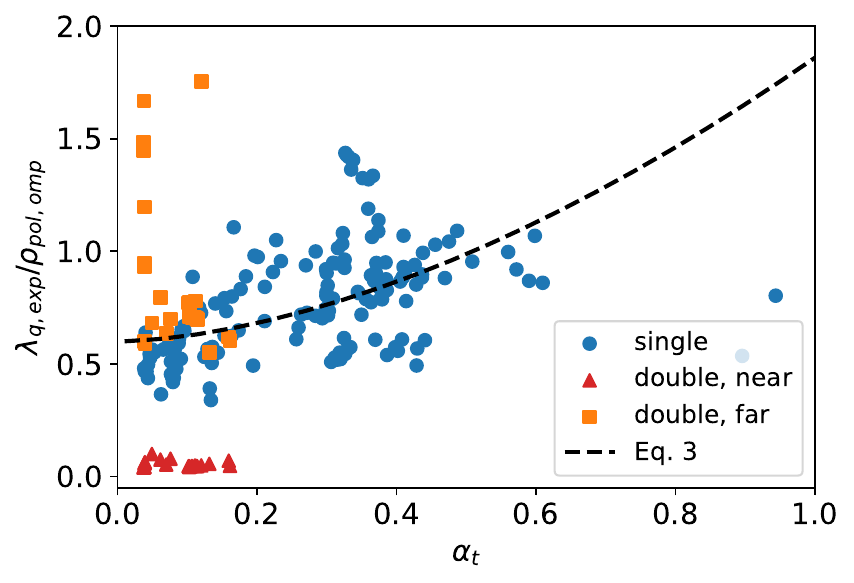}
    \caption{Comparison of measured scrape-off layer width against the ASDEX turbulence control parameter $\alpha_t$ scaling. The wide branch consisting of $\lqS$ and $\lqF$ follows the regression well. The onset of the narrow branch occurs at low values of $\alpha_t < 0.2$.}
    \label{fig:alphat}
\end{figure}

Turbulence-induced broadening of SOL widths has been observed experimentally and in high-fidelity simulations \cite{chang2017gyrokinetic, eich2020turbulence, faitsch2021broadening, ernst2024broadening}. The work on ASDEX by Eich \textit{et al} defined a turbulence control parameter $\alpha_t \doteq q_{cyl}\nu_{e}^*/100$, where $\nu_e^*$ is the edge electron collisionality, that accounts for this broadening. Figure~\ref{fig:alphat} shows the ST40 measurements against $\alpha_t$, with the dashed line representing temperature decay width ($\lambda_{T, e}$) regression results from ASDEX converted to power decay width by assuming Spitzer-H\"arm conductivity \cite{eich2020turbulence, faitsch2020correlation}:
\begin{equation}
    \frac{\lambda_q}{\rho_{pol}} = \frac{2}{7} \frac{\lambda_{T_e}}{\rho_{pol}} = 0.6 (1 + 2.1\alpha_t^{1.7}).
\end{equation}
where $\rho_{pol}$ is the ion poloidal Larmor radius. 

Electron temperature and density measured near the separatrix are used to calculate $\alpha_t$, with $Z_{eff} = 2$ based on visible Bremsstrahlung measurements. The separatrix electron temperature $T_{e, sep}$ is between 75 and 100 eV, close to power balance estimates using $\lqF$. Exact validation of the separatrix position based on power balance arguments \cite{stangeby2000plasma, zhang2023reduced, hecko2023experimental, faitsch2020correlation} is not suitable here because the power sharing between the near and far SOL is unclear. IR frames in windows where TS profiles showed abrupt changes were removed due to the high associated uncertainty in separatrix quantities. 

%, which indicates that the bifurcation of near SOL widths occurs at low $\alpha_t$

It can be seen that the wide branch, once again consisting of $\lqS$ and $\lqF$, follows the $\alpha_t$ regression well in trend and absolute value, with a similar level of scatter as previous publications from ASDEX and JET \cite{eich2020turbulence, faitsch2020correlation}. The narrow branch is observed exclusively at low $\alpha_t$ values below 0.2. In fact, the turbulence control parameter $\alpha_t$ is the only effective discriminator that can divide the parameter space so that the onset of the narrow branch corresponds to a clear threshold value. The uncertainty in separatrix location will result in a systematic shift of the horizontal axis, but does not influence the qualitative behavior nor the conclusion.

%dividing the parameter space where only the wide branch is observed and the onset of the narrow branch with a threshold value.

\paragraph{Discussion.}
A likely explanation of the bifurcation into the narrow branch is therefore attributed to the suppression of electromagnetic turbulence and electron anomalous transport, and the subsequent formation of a transport barrier in the SOL. The reduced turbulent transport leads to sharp gradients in the near SOL as well as in the pedestal region, manifesting as high pedestal pressure, narrow $\lqN$, and low $\alpha_t$. In such cases, a strong radial electric field is necessarily created, which leads to $E\times B$ shear that further suppresses turbulent transport and reinforces the sharp gradient - similar to the non-linear process that leads to the formation of the pedestal transport barrier \cite{wagner1984development, beyer2005nonlinear}. A related explanation is the creation of non-ambipolar currents in the SOL, which has been reported to contribute to narrow features in the near SOL heat flux profiles of limited and diverted plasmas \cite{brida2020role, labit2016physics, dejarnac2015understanding}.

Similar narrow SOL widths that are 2-5 times below scalings have been reported at TCV \cite{maurizio2021h} and COMPASS \cite{hecko2023experimental}, at poloidal magnetic field similar to NSTX. The difference between the measured $\lqN$ across these devices was attributed to the difference in aspect ratio, with $A = 1.25$ for NSTX, $A = 2.4$ for COMPASS, and $A = 3.5$ for TCV. However, since a narrow SOL width is also observed in ST40, which has aspect ratio similar to NSTX at $A = 1.5$ but pressure, fields, and input power similar to COMPASS, this suggests that the bifurcation is actually unrelated to aspect ratio. In fact, at $B_{p, omp} \sim 0.2$~T, the ST40 wide branch agrees with the values previously reported at NSTX, while the narrow branch agrees with the values recently reported at COMPASS \cite{eich2013scaling, hecko2023experimental}. 

Interestingly, the width of the narrow branch observed at ST40, clustering around 0.4~mm, is approximately equal to the ion Larmor radius at the OMP, which is between 0.4 - 0.7~mm assuming $T_{i, sep} = T_{e, sep} = 50 - 70$~eV. The $1/B_{t}$ dependency of the total Larmor radius can then roughly explain the sequence of $\lqN$ measurements going from COMPASS and ST40 (0.4-0.8~mm at 1.5~T), to TCV (2mm at 1~T), and then to NSTX and MAST (4 - 8~mm at 0.5~T). Similar dependency on $B_t$ is not observed in the wide branch both within the ST40 dataset or the previously published ITPA results \cite{eich2013scaling}.

%(albeit observed only transiently at ST40)

Further investigations both in experiments and theory are needed to understand the bifurcation of near SOL widths observed at ST40, and possibly COMPASS and TCV - all three of which were not included in the ITPA dataset. These three consecutive reports of near SOL widths far below scalings are sobering, highlighting the limitations of regression laws and calling for deeper physics-based understanding. Future experimental endeavours on ST40 will focus on further characterization of the SOL, including OMP profile measurements with an upgraded TS system to resolve $T_e < 50$~eV, re-calibrated IR camera that measures up to 200 degree Celsius, and profile and fluctuation measurements from the recently commissioned Langmuir probe arrays on both the upper and lower outer divertors. Dedicated experimental  scans in magnetic field, current, input power, and plasma density / fueling will also be performed to evaluate parametric dependencies and the controllability of the two branches. On the theoretical front, understanding the mechanism of magnetic fluctuation suppression and the  steepening of gradients in the near SOL can shed light on the nature of this narrow transport channel, and also help predict the power sharing between the near and far SOL. Such understanding will be crucial for predicting both the peak parallel heat flux and the shape of the divertor footprint, both of which are critical for designing mitigation and heat flux handling solutions for the next-generation tokamak reactors.

\paragraph{Acknowledgement.} The submitted manuscript has been co-authored by a contractor of the U.S. Government under contract DE-AC05-00OR22725. Accordingly, the U.S. Government retains a nonexclusive, royalty-free license to publish or reproduce the published form of this contribution, or allow others to do so, for U.S. Government purposes. The experimental data supporting the findings in this work may be made available upon reasonable request to the author.

\bibliography{biblio.bib}

%apsrev4-2.bst 2019-01-14 (MD) hand-edited version of apsrev4-1.bst
%Control: key (0)
%Control: author (8) initials jnrlst
%Control: editor formatted (1) identically to author
%Control: production of article title (0) allowed
%Control: page (0) single
%Control: year (1) truncated
%Control: production of eprint (0) enabled
\begin{thebibliography}{32}%
\makeatletter
\providecommand \@ifxundefined [1]{%
 \@ifx{#1\undefined}
}%
\providecommand \@ifnum [1]{%
 \ifnum #1\expandafter \@firstoftwo
 \else \expandafter \@secondoftwo
 \fi
}%
\providecommand \@ifx [1]{%
 \ifx #1\expandafter \@firstoftwo
 \else \expandafter \@secondoftwo
 \fi
}%
\providecommand \natexlab [1]{#1}%
\providecommand \enquote  [1]{``#1''}%
\providecommand \bibnamefont  [1]{#1}%
\providecommand \bibfnamefont [1]{#1}%
\providecommand \citenamefont [1]{#1}%
\providecommand \href@noop [0]{\@secondoftwo}%
\providecommand \href [0]{\begingroup \@sanitize@url \@href}%
\providecommand \@href[1]{\@@startlink{#1}\@@href}%
\providecommand \@@href[1]{\endgroup#1\@@endlink}%
\providecommand \@sanitize@url [0]{\catcode `\\12\catcode `\$12\catcode `\&12\catcode `\#12\catcode `\^12\catcode `\_12\catcode `\%12\relax}%
\providecommand \@@startlink[1]{}%
\providecommand \@@endlink[0]{}%
\providecommand \url  [0]{\begingroup\@sanitize@url \@url }%
\providecommand \@url [1]{\endgroup\@href {#1}{\urlprefix }}%
\providecommand \urlprefix  [0]{URL }%
\providecommand \Eprint [0]{\href }%
\providecommand \doibase [0]{https://doi.org/}%
\providecommand \selectlanguage [0]{\@gobble}%
\providecommand \bibinfo  [0]{\@secondoftwo}%
\providecommand \bibfield  [0]{\@secondoftwo}%
\providecommand \translation [1]{[#1]}%
\providecommand \BibitemOpen [0]{}%
\providecommand \bibitemStop [0]{}%
\providecommand \bibitemNoStop [0]{.\EOS\space}%
\providecommand \EOS [0]{\spacefactor3000\relax}%
\providecommand \BibitemShut  [1]{\csname bibitem#1\endcsname}%
\let\auto@bib@innerbib\@empty
%</preamble>
\bibitem [{\citenamefont {Antar}\ \emph {et~al.}(2010)\citenamefont {Antar}, \citenamefont {Assas}, \citenamefont {Bobkov}, \citenamefont {Noterdaeme}, \citenamefont {Wolfrum}, \citenamefont {Herrmann},\ and\ \citenamefont {Rohde}}]{antar2010convective}%
  \BibitemOpen
  \bibfield  {author} {\bibinfo {author} {\bibfnamefont {G.}~\bibnamefont {Antar}}, \bibinfo {author} {\bibfnamefont {S.}~\bibnamefont {Assas}}, \bibinfo {author} {\bibfnamefont {V.}~\bibnamefont {Bobkov}}, \bibinfo {author} {\bibfnamefont {J.-M.}\ \bibnamefont {Noterdaeme}}, \bibinfo {author} {\bibfnamefont {E.}~\bibnamefont {Wolfrum}}, \bibinfo {author} {\bibfnamefont {A.}~\bibnamefont {Herrmann}},\ and\ \bibinfo {author} {\bibfnamefont {V.}~\bibnamefont {Rohde}},\ }\bibfield  {title} {\bibinfo {title} {Convective transport suppression in the scrape-off layer using ion cyclotron resonance heating on the {ASDEX Upgrade} tokamak},\ }\href {https://doi.org/10.1103/PhysRevLett.105.165001} {\bibfield  {journal} {\bibinfo  {journal} {Phys. Rev. Lett.}\ }\textbf {\bibinfo {volume} {105}},\ \bibinfo {pages} {165001} (\bibinfo {year} {2010})}\BibitemShut {NoStop}%
\bibitem [{\citenamefont {Lunt}\ \emph {et~al.}(2023)\citenamefont {Lunt}, \citenamefont {Bernert}, \citenamefont {Brida}, \citenamefont {David}, \citenamefont {Faitsch}, \citenamefont {Pan}, \citenamefont {Stieglitz}, \citenamefont {Stroth}, \citenamefont {Redl},\ and\ \citenamefont {{ASDEX Upgrade Team}}}]{lunt2023compact}%
  \BibitemOpen
  \bibfield  {author} {\bibinfo {author} {\bibfnamefont {T.}~\bibnamefont {Lunt}}, \bibinfo {author} {\bibfnamefont {M.}~\bibnamefont {Bernert}}, \bibinfo {author} {\bibfnamefont {D.}~\bibnamefont {Brida}}, \bibinfo {author} {\bibfnamefont {P.}~\bibnamefont {David}}, \bibinfo {author} {\bibfnamefont {M.}~\bibnamefont {Faitsch}}, \bibinfo {author} {\bibfnamefont {O.}~\bibnamefont {Pan}}, \bibinfo {author} {\bibfnamefont {D.}~\bibnamefont {Stieglitz}}, \bibinfo {author} {\bibfnamefont {U.}~\bibnamefont {Stroth}}, \bibinfo {author} {\bibfnamefont {A.}~\bibnamefont {Redl}},\ and\ \bibinfo {author} {\bibnamefont {{ASDEX Upgrade Team}}},\ }\bibfield  {title} {\bibinfo {title} {Compact radiative divertor experiments at {ASDEX Upgrade} and their consequences for a reactor},\ }\href {https://doi.org/10.1103/PhysRevLett.130.145102} {\bibfield  {journal} {\bibinfo  {journal} {Phys. Rev. Lett.}\ }\textbf {\bibinfo {volume} {130}},\ \bibinfo {pages} {145102} (\bibinfo {year} {2023})}\BibitemShut {NoStop}%
\bibitem [{\citenamefont {Nelson}\ \emph {et~al.}(2023)\citenamefont {Nelson}, \citenamefont {Schmitz}, \citenamefont {Paz-Soldan}, \citenamefont {Thome}, \citenamefont {Cote}, \citenamefont {Leuthold}, \citenamefont {Scotti}, \citenamefont {Austin}, \citenamefont {Hyatt},\ and\ \citenamefont {Osborne}}]{nelson2023robust}%
  \BibitemOpen
  \bibfield  {author} {\bibinfo {author} {\bibfnamefont {A.}~\bibnamefont {Nelson}}, \bibinfo {author} {\bibfnamefont {L.}~\bibnamefont {Schmitz}}, \bibinfo {author} {\bibfnamefont {C.}~\bibnamefont {Paz-Soldan}}, \bibinfo {author} {\bibfnamefont {K.}~\bibnamefont {Thome}}, \bibinfo {author} {\bibfnamefont {T.}~\bibnamefont {Cote}}, \bibinfo {author} {\bibfnamefont {N.}~\bibnamefont {Leuthold}}, \bibinfo {author} {\bibfnamefont {F.}~\bibnamefont {Scotti}}, \bibinfo {author} {\bibfnamefont {M.}~\bibnamefont {Austin}}, \bibinfo {author} {\bibfnamefont {A.}~\bibnamefont {Hyatt}},\ and\ \bibinfo {author} {\bibfnamefont {T.}~\bibnamefont {Osborne}},\ }\bibfield  {title} {\bibinfo {title} {Robust avoidance of edge-localized modes alongside gradient formation in the negative triangularity tokamak edge},\ }\href {https://doi.org/10.1103/PhysRevLett.131.195101} {\bibfield  {journal} {\bibinfo  {journal} {Phys. Rev. Lett.}\ }\textbf {\bibinfo {volume} {131}},\ \bibinfo {pages} {195101} (\bibinfo {year}
  {2023})}\BibitemShut {NoStop}%
\bibitem [{\citenamefont {Fasoli}(2023)}]{fasoli2023essay}%
  \BibitemOpen
  \bibfield  {author} {\bibinfo {author} {\bibfnamefont {A.}~\bibnamefont {Fasoli}},\ }\bibfield  {title} {\bibinfo {title} {Essay: Overcoming the obstacles to a magnetic fusion power plant},\ }\href {https://doi.org/10.1103/PhysRevLett.130.220001} {\bibfield  {journal} {\bibinfo  {journal} {Phys. Rev. Lett.}\ }\textbf {\bibinfo {volume} {130}},\ \bibinfo {pages} {220001} (\bibinfo {year} {2023})}\BibitemShut {NoStop}%
\bibitem [{\citenamefont {Perkins}\ \emph {et~al.}(2012)\citenamefont {Perkins}, \citenamefont {Hosea}, \citenamefont {Kramer}, \citenamefont {Ahn}, \citenamefont {Bell}, \citenamefont {Diallo}, \citenamefont {Gerhardt}, \citenamefont {Gray}, \citenamefont {Green}, \citenamefont {Jaeger} \emph {et~al.}}]{perkins2012high}%
  \BibitemOpen
  \bibfield  {author} {\bibinfo {author} {\bibfnamefont {R.}~\bibnamefont {Perkins}}, \bibinfo {author} {\bibfnamefont {J.}~\bibnamefont {Hosea}}, \bibinfo {author} {\bibfnamefont {G.}~\bibnamefont {Kramer}}, \bibinfo {author} {\bibfnamefont {J.-W.}\ \bibnamefont {Ahn}}, \bibinfo {author} {\bibfnamefont {R.}~\bibnamefont {Bell}}, \bibinfo {author} {\bibfnamefont {A.}~\bibnamefont {Diallo}}, \bibinfo {author} {\bibfnamefont {S.}~\bibnamefont {Gerhardt}}, \bibinfo {author} {\bibfnamefont {T.}~\bibnamefont {Gray}}, \bibinfo {author} {\bibfnamefont {D.~L.}\ \bibnamefont {Green}}, \bibinfo {author} {\bibfnamefont {E.~F.}\ \bibnamefont {Jaeger}}, \emph {et~al.},\ }\bibfield  {title} {\bibinfo {title} {High-harmonic fast-wave power flow along magnetic field lines in the scrape-off layer of {NSTX}},\ }\href {https://doi.org/10.1103/PhysRevLett.109.045001} {\bibfield  {journal} {\bibinfo  {journal} {Phys. Rev. Lett.}\ }\textbf {\bibinfo {volume} {109}},\ \bibinfo {pages} {045001} (\bibinfo {year} {2012})}\BibitemShut
  {NoStop}%
\bibitem [{\citenamefont {Suttrop}\ \emph {et~al.}(2011)\citenamefont {Suttrop}, \citenamefont {Eich}, \citenamefont {Fuchs}, \citenamefont {G{\"u}nter}, \citenamefont {Janzer}, \citenamefont {Herrmann}, \citenamefont {Kallenbach}, \citenamefont {Lang}, \citenamefont {Lunt}, \citenamefont {Maraschek} \emph {et~al.}}]{suttrop2011first}%
  \BibitemOpen
  \bibfield  {author} {\bibinfo {author} {\bibfnamefont {W.}~\bibnamefont {Suttrop}}, \bibinfo {author} {\bibfnamefont {T.}~\bibnamefont {Eich}}, \bibinfo {author} {\bibfnamefont {J.}~\bibnamefont {Fuchs}}, \bibinfo {author} {\bibfnamefont {S.}~\bibnamefont {G{\"u}nter}}, \bibinfo {author} {\bibfnamefont {A.}~\bibnamefont {Janzer}}, \bibinfo {author} {\bibfnamefont {A.}~\bibnamefont {Herrmann}}, \bibinfo {author} {\bibfnamefont {A.}~\bibnamefont {Kallenbach}}, \bibinfo {author} {\bibfnamefont {P.}~\bibnamefont {Lang}}, \bibinfo {author} {\bibfnamefont {T.}~\bibnamefont {Lunt}}, \bibinfo {author} {\bibfnamefont {M.}~\bibnamefont {Maraschek}}, \emph {et~al.},\ }\bibfield  {title} {\bibinfo {title} {First observation of edge localized modes mitigation with resonant and nonresonant magnetic perturbations in {ASDEX Upgrade}},\ }\href {https://doi.org/10.1103/PhysRevLett.106.225004} {\bibfield  {journal} {\bibinfo  {journal} {Phys. Rev. Lett.}\ }\textbf {\bibinfo {volume} {106}},\ \bibinfo {pages} {225004}
  (\bibinfo {year} {2011})}\BibitemShut {NoStop}%
\bibitem [{\citenamefont {Eich}\ \emph {et~al.}(2013)\citenamefont {Eich}, \citenamefont {Leonard}, \citenamefont {Pitts}, \citenamefont {Fundamenski}, \citenamefont {Goldston}, \citenamefont {Gray}, \citenamefont {Herrmann}, \citenamefont {Kirk}, \citenamefont {Kallenbach}, \citenamefont {Kardaun} \emph {et~al.}}]{eich2013scaling}%
  \BibitemOpen
  \bibfield  {author} {\bibinfo {author} {\bibfnamefont {T.}~\bibnamefont {Eich}}, \bibinfo {author} {\bibfnamefont {A.}~\bibnamefont {Leonard}}, \bibinfo {author} {\bibfnamefont {R.}~\bibnamefont {Pitts}}, \bibinfo {author} {\bibfnamefont {W.}~\bibnamefont {Fundamenski}}, \bibinfo {author} {\bibfnamefont {R.~J.}\ \bibnamefont {Goldston}}, \bibinfo {author} {\bibfnamefont {T.}~\bibnamefont {Gray}}, \bibinfo {author} {\bibfnamefont {A.}~\bibnamefont {Herrmann}}, \bibinfo {author} {\bibfnamefont {A.}~\bibnamefont {Kirk}}, \bibinfo {author} {\bibfnamefont {A.}~\bibnamefont {Kallenbach}}, \bibinfo {author} {\bibfnamefont {O.}~\bibnamefont {Kardaun}}, \emph {et~al.},\ }\bibfield  {title} {\bibinfo {title} {Scaling of the tokamak near the scrape-off layer {H-mode} power width and implications for {ITER}},\ }\href {https://doi.org/10.1088/0029-5515/53/9/093031} {\bibfield  {journal} {\bibinfo  {journal} {Nucl. Fusion}\ }\textbf {\bibinfo {volume} {53}},\ \bibinfo {pages} {093031} (\bibinfo {year}
  {2013})}\BibitemShut {NoStop}%
\bibitem [{\citenamefont {Silvagni}\ \emph {et~al.}(2020)\citenamefont {Silvagni}, \citenamefont {Eich}, \citenamefont {Faitsch}, \citenamefont {Happel}, \citenamefont {Sieglin}, \citenamefont {David}, \citenamefont {Nille}, \citenamefont {Gil}, \citenamefont {Stroth}, \citenamefont {{ASDEX Upgrade Team}},\ and\ \citenamefont {{the EUROfusion MST1 team}}}]{silvagni2020scrape}%
  \BibitemOpen
  \bibfield  {author} {\bibinfo {author} {\bibfnamefont {D.}~\bibnamefont {Silvagni}}, \bibinfo {author} {\bibfnamefont {T.}~\bibnamefont {Eich}}, \bibinfo {author} {\bibfnamefont {M.}~\bibnamefont {Faitsch}}, \bibinfo {author} {\bibfnamefont {T.}~\bibnamefont {Happel}}, \bibinfo {author} {\bibfnamefont {B.}~\bibnamefont {Sieglin}}, \bibinfo {author} {\bibfnamefont {P.}~\bibnamefont {David}}, \bibinfo {author} {\bibfnamefont {D.}~\bibnamefont {Nille}}, \bibinfo {author} {\bibfnamefont {L.}~\bibnamefont {Gil}}, \bibinfo {author} {\bibfnamefont {U.}~\bibnamefont {Stroth}}, \bibinfo {author} {\bibnamefont {{ASDEX Upgrade Team}}},\ and\ \bibinfo {author} {\bibnamefont {{the EUROfusion MST1 team}}},\ }\bibfield  {title} {\bibinfo {title} {Scrape-off layer {(SOL)} power width scaling and correlation between {SOL} and pedestal gradients across {L, I and H-mode} plasmas at {ASDEX Upgrade}},\ }\href {https://doi.org/10.1088/1361-6587/ab74e8} {\bibfield  {journal} {\bibinfo  {journal} {Plasma Phys. and Control.
  Fusion}\ }\textbf {\bibinfo {volume} {62}},\ \bibinfo {pages} {045015} (\bibinfo {year} {2020})}\BibitemShut {NoStop}%
\bibitem [{\citenamefont {Ballinger}\ \emph {et~al.}(2022)\citenamefont {Ballinger}, \citenamefont {Brunner}, \citenamefont {Hubbard}, \citenamefont {Hughes}, \citenamefont {Kuang}, \citenamefont {LaBombard}, \citenamefont {Terry},\ and\ \citenamefont {White}}]{ballinger2022dependence}%
  \BibitemOpen
  \bibfield  {author} {\bibinfo {author} {\bibfnamefont {S.}~\bibnamefont {Ballinger}}, \bibinfo {author} {\bibfnamefont {D.}~\bibnamefont {Brunner}}, \bibinfo {author} {\bibfnamefont {A.}~\bibnamefont {Hubbard}}, \bibinfo {author} {\bibfnamefont {J.}~\bibnamefont {Hughes}}, \bibinfo {author} {\bibfnamefont {A.}~\bibnamefont {Kuang}}, \bibinfo {author} {\bibfnamefont {B.}~\bibnamefont {LaBombard}}, \bibinfo {author} {\bibfnamefont {J.}~\bibnamefont {Terry}},\ and\ \bibinfo {author} {\bibfnamefont {A.}~\bibnamefont {White}},\ }\bibfield  {title} {\bibinfo {title} {Dependence of the boundary heat flux width on core and edge profiles in {Alcator C-Mod}},\ }\href {https://doi.org/10.1088/1741-4326/ac637c} {\bibfield  {journal} {\bibinfo  {journal} {Nucl. Fusion}\ }\textbf {\bibinfo {volume} {62}},\ \bibinfo {pages} {076020} (\bibinfo {year} {2022})}\BibitemShut {NoStop}%
\bibitem [{\citenamefont {Faitsch}\ \emph {et~al.}(2020)\citenamefont {Faitsch}, \citenamefont {Eich}, \citenamefont {Sieglin},\ and\ \citenamefont {{JET Contributors}}}]{faitsch2020correlation}%
  \BibitemOpen
  \bibfield  {author} {\bibinfo {author} {\bibfnamefont {M.}~\bibnamefont {Faitsch}}, \bibinfo {author} {\bibfnamefont {T.}~\bibnamefont {Eich}}, \bibinfo {author} {\bibfnamefont {B.}~\bibnamefont {Sieglin}},\ and\ \bibinfo {author} {\bibnamefont {{JET Contributors}}},\ }\bibfield  {title} {\bibinfo {title} {Correlation between near scrape-off layer power fall-off length and confinement properties in {JET} operated with carbon and {ITER}-like wall},\ }\href {https://doi.org/10.1088/1361-6587/ab9073} {\bibfield  {journal} {\bibinfo  {journal} {Plasma Phys. Control. Fusion}\ }\textbf {\bibinfo {volume} {62}},\ \bibinfo {pages} {085004} (\bibinfo {year} {2020})}\BibitemShut {NoStop}%
\bibitem [{\citenamefont {Goldston}(2011)}]{goldston2011heuristic}%
  \BibitemOpen
  \bibfield  {author} {\bibinfo {author} {\bibfnamefont {R.~J.}\ \bibnamefont {Goldston}},\ }\bibfield  {title} {\bibinfo {title} {Heuristic drift-based model of the power scrape-off width in low-gas-puff {H-mode} tokamaks},\ }\href {https://doi.org/10.1088/0029-5515/52/1/013009} {\bibfield  {journal} {\bibinfo  {journal} {Nucl. Fusion}\ }\textbf {\bibinfo {volume} {52}},\ \bibinfo {pages} {013009} (\bibinfo {year} {2011})}\BibitemShut {NoStop}%
\bibitem [{\citenamefont {Eich}\ \emph {et~al.}(2011)\citenamefont {Eich}, \citenamefont {Sieglin}, \citenamefont {Scarabosio}, \citenamefont {Fundamenski}, \citenamefont {Goldston}, \citenamefont {Herrmann},\ and\ \citenamefont {{ASDEX Upgrade Team}}}]{eich2011inter}%
  \BibitemOpen
  \bibfield  {author} {\bibinfo {author} {\bibfnamefont {T.}~\bibnamefont {Eich}}, \bibinfo {author} {\bibfnamefont {B.}~\bibnamefont {Sieglin}}, \bibinfo {author} {\bibfnamefont {A.}~\bibnamefont {Scarabosio}}, \bibinfo {author} {\bibfnamefont {W.}~\bibnamefont {Fundamenski}}, \bibinfo {author} {\bibfnamefont {R.~J.}\ \bibnamefont {Goldston}}, \bibinfo {author} {\bibfnamefont {A.}~\bibnamefont {Herrmann}},\ and\ \bibinfo {author} {\bibnamefont {{ASDEX Upgrade Team}}},\ }\bibfield  {title} {\bibinfo {title} {Inter-{ELM} power decay length for {JET} and {ASDEX Upgrade}: measurement and comparison with heuristic drift-based model},\ }\href {https://doi.org/10.1103/PhysRevLett.107.215001} {\bibfield  {journal} {\bibinfo  {journal} {Phys. Rev. Lett.}\ }\textbf {\bibinfo {volume} {107}},\ \bibinfo {pages} {215001} (\bibinfo {year} {2011})}\BibitemShut {NoStop}%
\bibitem [{\citenamefont {Goldston}(2015)}]{goldston2015theoretical}%
  \BibitemOpen
  \bibfield  {author} {\bibinfo {author} {\bibfnamefont {R.}~\bibnamefont {Goldston}},\ }\bibfield  {title} {\bibinfo {title} {Theoretical aspects and practical implications of the heuristic drift {SOL} model},\ }\href {https://doi.org/10.1016/j.jnucmat.2014.10.080} {\bibfield  {journal} {\bibinfo  {journal} {J. Nucl. Mater.}\ }\textbf {\bibinfo {volume} {463}},\ \bibinfo {pages} {397} (\bibinfo {year} {2015})}\BibitemShut {NoStop}%
\bibitem [{\citenamefont {Chang}\ \emph {et~al.}(2017)\citenamefont {Chang}, \citenamefont {Ku}, \citenamefont {Loarte}, \citenamefont {Parail}, \citenamefont {Koechl}, \citenamefont {Romanelli}, \citenamefont {Maingi}, \citenamefont {Ahn}, \citenamefont {Gray}, \citenamefont {Hughes} \emph {et~al.}}]{chang2017gyrokinetic}%
  \BibitemOpen
  \bibfield  {author} {\bibinfo {author} {\bibfnamefont {C.~S.}\ \bibnamefont {Chang}}, \bibinfo {author} {\bibfnamefont {S.}~\bibnamefont {Ku}}, \bibinfo {author} {\bibfnamefont {A.}~\bibnamefont {Loarte}}, \bibinfo {author} {\bibfnamefont {V.}~\bibnamefont {Parail}}, \bibinfo {author} {\bibfnamefont {F.}~\bibnamefont {Koechl}}, \bibinfo {author} {\bibfnamefont {M.}~\bibnamefont {Romanelli}}, \bibinfo {author} {\bibfnamefont {R.}~\bibnamefont {Maingi}}, \bibinfo {author} {\bibfnamefont {J.-W.}\ \bibnamefont {Ahn}}, \bibinfo {author} {\bibfnamefont {T.}~\bibnamefont {Gray}}, \bibinfo {author} {\bibfnamefont {J.}~\bibnamefont {Hughes}}, \emph {et~al.},\ }\bibfield  {title} {\bibinfo {title} {Gyrokinetic projection of the divertor heat-flux width from present tokamaks to {ITER}},\ }\href {https://doi.org/10.1088/1741-4326/aa7efb} {\bibfield  {journal} {\bibinfo  {journal} {Nucl. Fusion}\ }\textbf {\bibinfo {volume} {57}},\ \bibinfo {pages} {116023} (\bibinfo {year} {2017})}\BibitemShut {NoStop}%
\bibitem [{\citenamefont {Eich}\ \emph {et~al.}(2020)\citenamefont {Eich}, \citenamefont {Manz}, \citenamefont {Goldston}, \citenamefont {Hennequin}, \citenamefont {David}, \citenamefont {Faitsch}, \citenamefont {Kurzan}, \citenamefont {Sieglin}, \citenamefont {Wolfrum}, \citenamefont {{ASDEX Upgrade Team}},\ and\ \citenamefont {{EUROfusion MST1 team}}}]{eich2020turbulence}%
  \BibitemOpen
  \bibfield  {author} {\bibinfo {author} {\bibfnamefont {T.}~\bibnamefont {Eich}}, \bibinfo {author} {\bibfnamefont {P.}~\bibnamefont {Manz}}, \bibinfo {author} {\bibfnamefont {R.}~\bibnamefont {Goldston}}, \bibinfo {author} {\bibfnamefont {P.}~\bibnamefont {Hennequin}}, \bibinfo {author} {\bibfnamefont {P.}~\bibnamefont {David}}, \bibinfo {author} {\bibfnamefont {M.}~\bibnamefont {Faitsch}}, \bibinfo {author} {\bibfnamefont {B.}~\bibnamefont {Kurzan}}, \bibinfo {author} {\bibfnamefont {B.}~\bibnamefont {Sieglin}}, \bibinfo {author} {\bibfnamefont {E.}~\bibnamefont {Wolfrum}}, \bibinfo {author} {\bibnamefont {{ASDEX Upgrade Team}}},\ and\ \bibinfo {author} {\bibnamefont {{EUROfusion MST1 team}}},\ }\bibfield  {title} {\bibinfo {title} {Turbulence driven widening of the near-{SOL} power width in {ASDEX Upgrade} {H-Mode} discharges},\ }\href {https://doi.org/10.1088/1741-4326/ab7a66} {\bibfield  {journal} {\bibinfo  {journal} {Nucl. Fusion}\ }\textbf {\bibinfo {volume} {60}},\ \bibinfo {pages} {056016}
  (\bibinfo {year} {2020})}\BibitemShut {NoStop}%
\bibitem [{\citenamefont {Ernst}\ \emph {et~al.}(pear)\citenamefont {Ernst}, \citenamefont {Bortolon}, \citenamefont {Chang}, \citenamefont {Ku}, \citenamefont {Scotti}, \citenamefont {Wang}, \citenamefont {Yan}, \citenamefont {Chen}, \citenamefont {Chrystal}, \citenamefont {Glass} \emph {et~al.}}]{ernst2024broadening}%
  \BibitemOpen
  \bibfield  {author} {\bibinfo {author} {\bibfnamefont {D.}~\bibnamefont {Ernst}}, \bibinfo {author} {\bibfnamefont {A.}~\bibnamefont {Bortolon}}, \bibinfo {author} {\bibfnamefont {C.}~\bibnamefont {Chang}}, \bibinfo {author} {\bibfnamefont {S.}~\bibnamefont {Ku}}, \bibinfo {author} {\bibfnamefont {F.}~\bibnamefont {Scotti}}, \bibinfo {author} {\bibfnamefont {H.}~\bibnamefont {Wang}}, \bibinfo {author} {\bibfnamefont {Z.}~\bibnamefont {Yan}}, \bibinfo {author} {\bibfnamefont {J.}~\bibnamefont {Chen}}, \bibinfo {author} {\bibfnamefont {C.}~\bibnamefont {Chrystal}}, \bibinfo {author} {\bibfnamefont {F.}~\bibnamefont {Glass}}, \emph {et~al.},\ }\bibfield  {title} {\bibinfo {title} {Broadening of the divertor heat flux profile in high confinement tokamak fusion plasmas with edge pedestals limited by turbulence in {DIII-D}},\ }\href@noop {} {\bibfield  {journal} {\bibinfo  {journal} {Phys. Rev. Lett.}\ } (\bibinfo {year} {to appear})}\BibitemShut {NoStop}%
\bibitem [{\citenamefont {Faitsch}\ \emph {et~al.}(2021)\citenamefont {Faitsch}, \citenamefont {Eich}, \citenamefont {Harrer}, \citenamefont {Wolfrum}, \citenamefont {Brida}, \citenamefont {David}, \citenamefont {Griener}, \citenamefont {Stroth}, \citenamefont {{ASDEX Upgrade Team}},\ and\ \citenamefont {{EUROfusion MST1 Team}}}]{faitsch2021broadening}%
  \BibitemOpen
  \bibfield  {author} {\bibinfo {author} {\bibfnamefont {M.}~\bibnamefont {Faitsch}}, \bibinfo {author} {\bibfnamefont {T.}~\bibnamefont {Eich}}, \bibinfo {author} {\bibfnamefont {G.}~\bibnamefont {Harrer}}, \bibinfo {author} {\bibfnamefont {E.}~\bibnamefont {Wolfrum}}, \bibinfo {author} {\bibfnamefont {D.}~\bibnamefont {Brida}}, \bibinfo {author} {\bibfnamefont {P.}~\bibnamefont {David}}, \bibinfo {author} {\bibfnamefont {M.}~\bibnamefont {Griener}}, \bibinfo {author} {\bibfnamefont {U.}~\bibnamefont {Stroth}}, \bibinfo {author} {\bibnamefont {{ASDEX Upgrade Team}}},\ and\ \bibinfo {author} {\bibnamefont {{EUROfusion MST1 Team}}},\ }\bibfield  {title} {\bibinfo {title} {Broadening of the power fall-off length in a high density, high confinement {H-mode} regime in {ASDEX Upgrade}},\ }\href {https://doi.org/10.1016/j.nme.2020.100890} {\bibfield  {journal} {\bibinfo  {journal} {Nucl. Mater. Energy}\ }\textbf {\bibinfo {volume} {26}},\ \bibinfo {pages} {100890} (\bibinfo {year} {2021})}\BibitemShut {NoStop}%
\bibitem [{\citenamefont {Maurizio}\ \emph {et~al.}(2021)\citenamefont {Maurizio}, \citenamefont {Duval}, \citenamefont {Labit}, \citenamefont {Reimerdes}, \citenamefont {Faitsch}, \citenamefont {Komm}, \citenamefont {Sheikh}, \citenamefont {Theiler}, \citenamefont {{TCV Team}},\ and\ \citenamefont {{EUROfusion MST1 team}}}]{maurizio2021h}%
  \BibitemOpen
  \bibfield  {author} {\bibinfo {author} {\bibfnamefont {R.}~\bibnamefont {Maurizio}}, \bibinfo {author} {\bibfnamefont {B.}~\bibnamefont {Duval}}, \bibinfo {author} {\bibfnamefont {B.}~\bibnamefont {Labit}}, \bibinfo {author} {\bibfnamefont {H.}~\bibnamefont {Reimerdes}}, \bibinfo {author} {\bibfnamefont {M.}~\bibnamefont {Faitsch}}, \bibinfo {author} {\bibfnamefont {M.}~\bibnamefont {Komm}}, \bibinfo {author} {\bibfnamefont {U.}~\bibnamefont {Sheikh}}, \bibinfo {author} {\bibfnamefont {C.}~\bibnamefont {Theiler}}, \bibinfo {author} {\bibnamefont {{TCV Team}}},\ and\ \bibinfo {author} {\bibnamefont {{EUROfusion MST1 team}}},\ }\bibfield  {title} {\bibinfo {title} {H-mode scrape-off layer power width in the {TCV} tokamak},\ }\href {https://doi.org/10.1088/1741-4326/abd147} {\bibfield  {journal} {\bibinfo  {journal} {Nucl. Fusion}\ }\textbf {\bibinfo {volume} {61}},\ \bibinfo {pages} {024003} (\bibinfo {year} {2021})}\BibitemShut {NoStop}%
\bibitem [{\citenamefont {Hecko}\ \emph {et~al.}(2023)\citenamefont {Hecko}, \citenamefont {Komm}, \citenamefont {Sos}, \citenamefont {Adamek}, \citenamefont {Bilkova}, \citenamefont {Bogar}, \citenamefont {Bohm}, \citenamefont {Jaulmes}, \citenamefont {Mysiura}, \citenamefont {Tomes} \emph {et~al.}}]{hecko2023experimental}%
  \BibitemOpen
  \bibfield  {author} {\bibinfo {author} {\bibfnamefont {J.}~\bibnamefont {Hecko}}, \bibinfo {author} {\bibfnamefont {M.}~\bibnamefont {Komm}}, \bibinfo {author} {\bibfnamefont {M.}~\bibnamefont {Sos}}, \bibinfo {author} {\bibfnamefont {J.}~\bibnamefont {Adamek}}, \bibinfo {author} {\bibfnamefont {P.}~\bibnamefont {Bilkova}}, \bibinfo {author} {\bibfnamefont {K.}~\bibnamefont {Bogar}}, \bibinfo {author} {\bibfnamefont {P.}~\bibnamefont {Bohm}}, \bibinfo {author} {\bibfnamefont {F.}~\bibnamefont {Jaulmes}}, \bibinfo {author} {\bibfnamefont {I.}~\bibnamefont {Mysiura}}, \bibinfo {author} {\bibfnamefont {M.}~\bibnamefont {Tomes}}, \emph {et~al.},\ }\bibfield  {title} {\bibinfo {title} {Experimental evidence of very short power decay lengths in h-mode discharges in the {COMPASS} tokamak},\ }\href {https://doi.org/10.1088/1361-6587/ad08f0} {\bibfield  {journal} {\bibinfo  {journal} {Plasma Phys. Control. Fusion}\ }\textbf {\bibinfo {volume} {66}},\ \bibinfo {pages} {015013} (\bibinfo {year} {2023})}\BibitemShut
  {NoStop}%
\bibitem [{\citenamefont {McNamara}\ \emph {et~al.}(2023)\citenamefont {McNamara}, \citenamefont {Asunta}, \citenamefont {Bland}, \citenamefont {Buxton}, \citenamefont {Colgan}, \citenamefont {Dnestrovskii}, \citenamefont {Gemmell}, \citenamefont {Gryaznevich}, \citenamefont {Hoffman}, \citenamefont {Janky} \emph {et~al.}}]{mcnamara2023achievement}%
  \BibitemOpen
  \bibfield  {author} {\bibinfo {author} {\bibfnamefont {S.}~\bibnamefont {McNamara}}, \bibinfo {author} {\bibfnamefont {O.}~\bibnamefont {Asunta}}, \bibinfo {author} {\bibfnamefont {J.}~\bibnamefont {Bland}}, \bibinfo {author} {\bibfnamefont {P.}~\bibnamefont {Buxton}}, \bibinfo {author} {\bibfnamefont {C.}~\bibnamefont {Colgan}}, \bibinfo {author} {\bibfnamefont {A.}~\bibnamefont {Dnestrovskii}}, \bibinfo {author} {\bibfnamefont {M.}~\bibnamefont {Gemmell}}, \bibinfo {author} {\bibfnamefont {M.}~\bibnamefont {Gryaznevich}}, \bibinfo {author} {\bibfnamefont {D.}~\bibnamefont {Hoffman}}, \bibinfo {author} {\bibfnamefont {F.}~\bibnamefont {Janky}}, \emph {et~al.},\ }\bibfield  {title} {\bibinfo {title} {Achievement of ion temperatures in excess of 100 million degrees {Kelvin} in the compact high-field spherical tokamak {ST40}},\ }\href {https://doi.org/10.1088/1741-4326/acbec8} {\bibfield  {journal} {\bibinfo  {journal} {Nucl. fusion}\ }\textbf {\bibinfo {volume} {63}},\ \bibinfo {pages} {054002} (\bibinfo
  {year} {2023})}\BibitemShut {NoStop}%
\bibitem [{\citenamefont {McNamara}\ \emph {et~al.}(view)\citenamefont {McNamara}, \citenamefont {Alieva}, \citenamefont {Anastopoulos~Tzanis}, \citenamefont {Asunta}, \citenamefont {Bland}, \citenamefont {Bohlin}, \citenamefont {Buxton}, \citenamefont {Colgan}, \citenamefont {Dnestrovskii}, \citenamefont {du~Toit} \emph {et~al.}}]{mcnamara2024overview}%
  \BibitemOpen
  \bibfield  {author} {\bibinfo {author} {\bibfnamefont {S.}~\bibnamefont {McNamara}}, \bibinfo {author} {\bibfnamefont {A.}~\bibnamefont {Alieva}}, \bibinfo {author} {\bibfnamefont {M.}~\bibnamefont {Anastopoulos~Tzanis}}, \bibinfo {author} {\bibfnamefont {O.}~\bibnamefont {Asunta}}, \bibinfo {author} {\bibfnamefont {J.}~\bibnamefont {Bland}}, \bibinfo {author} {\bibfnamefont {H.}~\bibnamefont {Bohlin}}, \bibinfo {author} {\bibfnamefont {P.}~\bibnamefont {Buxton}}, \bibinfo {author} {\bibfnamefont {C.}~\bibnamefont {Colgan}}, \bibinfo {author} {\bibfnamefont {A.}~\bibnamefont {Dnestrovskii}}, \bibinfo {author} {\bibfnamefont {E.}~\bibnamefont {du~Toit}}, \emph {et~al.},\ }\bibfield  {title} {\bibinfo {title} {Overview of recent results from the {ST40} compact high-field spherical tokamak},\ }\href@noop {} {\bibfield  {journal} {\bibinfo  {journal} {Nucl. Fusion}\ } (\bibinfo {year} {under review})}\BibitemShut {NoStop}%
\bibitem [{\citenamefont {Robinson}\ \emph {et~al.}(2024)\citenamefont {Robinson}, \citenamefont {Maartensson}, \citenamefont {Moscheni}, \citenamefont {Rengle}, \citenamefont {Jackson}, \citenamefont {Lee}, \citenamefont {Yildirim}, \citenamefont {Gray},\ and\ \citenamefont {Team}}]{PSI_MR}%
  \BibitemOpen
  \bibfield  {author} {\bibinfo {author} {\bibfnamefont {M.}~\bibnamefont {Robinson}}, \bibinfo {author} {\bibfnamefont {E.}~\bibnamefont {Maartensson}}, \bibinfo {author} {\bibfnamefont {M.}~\bibnamefont {Moscheni}}, \bibinfo {author} {\bibfnamefont {A.}~\bibnamefont {Rengle}}, \bibinfo {author} {\bibfnamefont {M.}~\bibnamefont {Jackson}}, \bibinfo {author} {\bibfnamefont {V.}~\bibnamefont {Lee}}, \bibinfo {author} {\bibfnamefont {E.}~\bibnamefont {Yildirim}}, \bibinfo {author} {\bibfnamefont {T.}~\bibnamefont {Gray}},\ and\ \bibinfo {author} {\bibfnamefont {t.}~\bibnamefont {Team}},\ }\href@noop {} {\bibinfo {title} {{ST40} tool for {IR} thermography: {FAHF}}} (\bibinfo {year} {2024}),\ \bibinfo {note} {{PSI26} Conference, Poster Contribution (P3-083)}\BibitemShut {NoStop}%
\bibitem [{\citenamefont {Moscheni}\ \emph {et~al.}(2024)\citenamefont {Moscheni}, \citenamefont {Marsden}, \citenamefont {Rengle}, \citenamefont {Robinson}, \citenamefont {Vekshina}, \citenamefont {Zhang}, \citenamefont {Chang}, \citenamefont {Gray}, \citenamefont {Hager}, \citenamefont {Janhunen}, \citenamefont {Ku}, \citenamefont {Lee}, \citenamefont {Lore}, \citenamefont {Maartensson}, \citenamefont {McNamara}, \citenamefont {Paradela~Pérez}, \citenamefont {Romanelli}, \citenamefont {Scarabosio}, \citenamefont {Tinacba}, \citenamefont {Unterberg},\ and\ \citenamefont {{the ST40 Team}}}]{PSI_MM}%
  \BibitemOpen
  \bibfield  {author} {\bibinfo {author} {\bibfnamefont {M.}~\bibnamefont {Moscheni}}, \bibinfo {author} {\bibfnamefont {C.}~\bibnamefont {Marsden}}, \bibinfo {author} {\bibfnamefont {A.}~\bibnamefont {Rengle}}, \bibinfo {author} {\bibfnamefont {M.}~\bibnamefont {Robinson}}, \bibinfo {author} {\bibfnamefont {E.}~\bibnamefont {Vekshina}}, \bibinfo {author} {\bibfnamefont {X.}~\bibnamefont {Zhang}}, \bibinfo {author} {\bibfnamefont {C.}~\bibnamefont {Chang}}, \bibinfo {author} {\bibfnamefont {T.}~\bibnamefont {Gray}}, \bibinfo {author} {\bibfnamefont {R.}~\bibnamefont {Hager}}, \bibinfo {author} {\bibfnamefont {S.}~\bibnamefont {Janhunen}}, \bibinfo {author} {\bibfnamefont {S.}~\bibnamefont {Ku}}, \bibinfo {author} {\bibfnamefont {V.}~\bibnamefont {Lee}}, \bibinfo {author} {\bibfnamefont {J.}~\bibnamefont {Lore}}, \bibinfo {author} {\bibfnamefont {E.}~\bibnamefont {Maartensson}}, \bibinfo {author} {\bibfnamefont {S.}~\bibnamefont {McNamara}}, \bibinfo {author} {\bibfnamefont {I.}~\bibnamefont
  {Paradela~Pérez}}, \bibinfo {author} {\bibfnamefont {M.}~\bibnamefont {Romanelli}}, \bibinfo {author} {\bibfnamefont {A.}~\bibnamefont {Scarabosio}}, \bibinfo {author} {\bibfnamefont {E.}~\bibnamefont {Tinacba}}, \bibinfo {author} {\bibfnamefont {E.}~\bibnamefont {Unterberg}},\ and\ \bibinfo {author} {\bibnamefont {{the ST40 Team}}},\ }\href@noop {} {\bibinfo {title} {Overview and preliminary assessment of divertor edge plasma experimental data in {ST40}}} (\bibinfo {year} {2024}),\ \bibinfo {note} {{PSI26} Conference, Poster Contribution (P1-095)}\BibitemShut {NoStop}%
\bibitem [{\citenamefont {Marsden}\ \emph {et~al.}(2024)\citenamefont {Marsden}, \citenamefont {Zhang}, \citenamefont {Moscheni}, \citenamefont {Gray}, \citenamefont {Vekshina}, \citenamefont {Rengle}, \citenamefont {Maartensson}, \citenamefont {Robinson}, \citenamefont {Scarabosio}, \citenamefont {Romanelli},\ and\ \citenamefont {{the ST40 team}}}]{PSI_CM}%
  \BibitemOpen
  \bibfield  {author} {\bibinfo {author} {\bibfnamefont {C.}~\bibnamefont {Marsden}}, \bibinfo {author} {\bibfnamefont {X.}~\bibnamefont {Zhang}}, \bibinfo {author} {\bibfnamefont {M.}~\bibnamefont {Moscheni}}, \bibinfo {author} {\bibfnamefont {T.}~\bibnamefont {Gray}}, \bibinfo {author} {\bibfnamefont {E.}~\bibnamefont {Vekshina}}, \bibinfo {author} {\bibfnamefont {S.}~\bibnamefont {Rengle}, \bibfnamefont {A.and~Janhunen}}, \bibinfo {author} {\bibfnamefont {E.}~\bibnamefont {Maartensson}}, \bibinfo {author} {\bibfnamefont {M.}~\bibnamefont {Robinson}}, \bibinfo {author} {\bibfnamefont {A.}~\bibnamefont {Scarabosio}}, \bibinfo {author} {\bibfnamefont {M.}~\bibnamefont {Romanelli}},\ and\ \bibinfo {author} {\bibnamefont {{the ST40 team}}},\ }\href@noop {} {\bibinfo {title} {Inferring the scrape-off layer heat flux width in {ST40} using the infra-red investigative analysis toolchain – {IRRITANT}}} (\bibinfo {year} {2024}),\ \bibinfo {note} {{PSI26} Conference, Poster Contribution (P3-076)}\BibitemShut
  {NoStop}%
\bibitem [{\citenamefont {Brunner}\ \emph {et~al.}(2018)\citenamefont {Brunner}, \citenamefont {LaBombard}, \citenamefont {Kuang},\ and\ \citenamefont {Terry}}]{brunner2018high}%
  \BibitemOpen
  \bibfield  {author} {\bibinfo {author} {\bibfnamefont {D.}~\bibnamefont {Brunner}}, \bibinfo {author} {\bibfnamefont {B.}~\bibnamefont {LaBombard}}, \bibinfo {author} {\bibfnamefont {A.}~\bibnamefont {Kuang}},\ and\ \bibinfo {author} {\bibfnamefont {J.}~\bibnamefont {Terry}},\ }\bibfield  {title} {\bibinfo {title} {High-resolution heat flux width measurements at reactor-level magnetic fields and observation of a unified width scaling across confinement regimes in the {Alcator C-Mod} tokamak},\ }\href {https://doi.org/10.1088/1741-4326/aad0d6} {\bibfield  {journal} {\bibinfo  {journal} {Nucl. Fusion}\ }\textbf {\bibinfo {volume} {58}},\ \bibinfo {pages} {094002} (\bibinfo {year} {2018})}\BibitemShut {NoStop}%
\bibitem [{\citenamefont {Stangeby}(2000)}]{stangeby2000plasma}%
  \BibitemOpen
  \bibfield  {author} {\bibinfo {author} {\bibfnamefont {P.~C.}\ \bibnamefont {Stangeby}},\ }\href {https://doi.org/10.1201/9780367801489} {\emph {\bibinfo {title} {The plasma boundary of magnetic fusion devices}}}\ (\bibinfo  {publisher} {New York: Taylor and Francis},\ \bibinfo {year} {2000})\BibitemShut {NoStop}%
\bibitem [{\citenamefont {Zhang}\ \emph {et~al.}(2023)\citenamefont {Zhang}, \citenamefont {Poli}, \citenamefont {Emdee}, \citenamefont {Podest{\`a}},\ and\ \citenamefont {{NSTX-U Team}}}]{zhang2023reduced}%
  \BibitemOpen
  \bibfield  {author} {\bibinfo {author} {\bibfnamefont {X.}~\bibnamefont {Zhang}}, \bibinfo {author} {\bibfnamefont {F.}~\bibnamefont {Poli}}, \bibinfo {author} {\bibfnamefont {E.}~\bibnamefont {Emdee}}, \bibinfo {author} {\bibfnamefont {M.}~\bibnamefont {Podest{\`a}}},\ and\ \bibinfo {author} {\bibnamefont {{NSTX-U Team}}},\ }\bibfield  {title} {\bibinfo {title} {Reduced physics model of the tokamak scrape-off-layer for pulse design},\ }\href {https://doi.org/10.1016/j.nme.2022.101354} {\bibfield  {journal} {\bibinfo  {journal} {Nucl. Mater. Energy}\ }\textbf {\bibinfo {volume} {34}},\ \bibinfo {pages} {101354} (\bibinfo {year} {2023})}\BibitemShut {NoStop}%
\bibitem [{\citenamefont {Wagner}\ \emph {et~al.}(1984)\citenamefont {Wagner}, \citenamefont {Fussmann}, \citenamefont {Grave}, \citenamefont {Keilhacker}, \citenamefont {Kornherr}, \citenamefont {Lackner}, \citenamefont {McCormick}, \citenamefont {M{\"u}ller}, \citenamefont {St{\"a}bler}, \citenamefont {Becker} \emph {et~al.}}]{wagner1984development}%
  \BibitemOpen
  \bibfield  {author} {\bibinfo {author} {\bibfnamefont {F.}~\bibnamefont {Wagner}}, \bibinfo {author} {\bibfnamefont {G.}~\bibnamefont {Fussmann}}, \bibinfo {author} {\bibfnamefont {T.}~\bibnamefont {Grave}}, \bibinfo {author} {\bibfnamefont {M.}~\bibnamefont {Keilhacker}}, \bibinfo {author} {\bibfnamefont {M.}~\bibnamefont {Kornherr}}, \bibinfo {author} {\bibfnamefont {K.}~\bibnamefont {Lackner}}, \bibinfo {author} {\bibfnamefont {K.}~\bibnamefont {McCormick}}, \bibinfo {author} {\bibfnamefont {E.}~\bibnamefont {M{\"u}ller}}, \bibinfo {author} {\bibfnamefont {A.}~\bibnamefont {St{\"a}bler}}, \bibinfo {author} {\bibfnamefont {G.}~\bibnamefont {Becker}}, \emph {et~al.},\ }\bibfield  {title} {\bibinfo {title} {Development of an edge transport barrier at the {H-mode} transition of {ASDEX}},\ }\href {https://doi.org/10.1103/PhysRevLett.53.1453} {\bibfield  {journal} {\bibinfo  {journal} {Phys. Rev. Lett.}\ }\textbf {\bibinfo {volume} {53}},\ \bibinfo {pages} {1453} (\bibinfo {year} {1984})}\BibitemShut {NoStop}%
\bibitem [{\citenamefont {Beyer}\ \emph {et~al.}(2005)\citenamefont {Beyer}, \citenamefont {Benkadda}, \citenamefont {Fuhr-Chaudier}, \citenamefont {Garbet}, \citenamefont {Ghendrih},\ and\ \citenamefont {Sarazin}}]{beyer2005nonlinear}%
  \BibitemOpen
  \bibfield  {author} {\bibinfo {author} {\bibfnamefont {P.}~\bibnamefont {Beyer}}, \bibinfo {author} {\bibfnamefont {S.}~\bibnamefont {Benkadda}}, \bibinfo {author} {\bibfnamefont {G.}~\bibnamefont {Fuhr-Chaudier}}, \bibinfo {author} {\bibfnamefont {X.}~\bibnamefont {Garbet}}, \bibinfo {author} {\bibfnamefont {P.}~\bibnamefont {Ghendrih}},\ and\ \bibinfo {author} {\bibfnamefont {Y.}~\bibnamefont {Sarazin}},\ }\bibfield  {title} {\bibinfo {title} {Nonlinear dynamics of transport barrier relaxations in tokamak edge plasmas},\ }\href {https://doi.org/10.1103/PhysRevLett.94.105001} {\bibfield  {journal} {\bibinfo  {journal} {Phys. Rev. Lett.}\ }\textbf {\bibinfo {volume} {94}},\ \bibinfo {pages} {105001} (\bibinfo {year} {2005})}\BibitemShut {NoStop}%
\bibitem [{\citenamefont {Brida}\ \emph {et~al.}(2020)\citenamefont {Brida}, \citenamefont {Silvagni}, \citenamefont {Eich}, \citenamefont {Faitsch}, \citenamefont {McCarthy}, \citenamefont {{ASDEX Upgrade Team}}, \citenamefont {{MST1 Team}} \emph {et~al.}}]{brida2020role}%
  \BibitemOpen
  \bibfield  {author} {\bibinfo {author} {\bibfnamefont {D.}~\bibnamefont {Brida}}, \bibinfo {author} {\bibfnamefont {D.}~\bibnamefont {Silvagni}}, \bibinfo {author} {\bibfnamefont {T.}~\bibnamefont {Eich}}, \bibinfo {author} {\bibfnamefont {M.}~\bibnamefont {Faitsch}}, \bibinfo {author} {\bibfnamefont {P.}~\bibnamefont {McCarthy}}, \bibinfo {author} {\bibnamefont {{ASDEX Upgrade Team}}}, \bibinfo {author} {\bibnamefont {{MST1 Team}}}, \emph {et~al.},\ }\bibfield  {title} {\bibinfo {title} {Role of electric currents for the sol and divertor target heat fluxes in {ASDEX} upgrade},\ }\href {https://iopscience.iop.org/article/10.1088/1361-6587/aba8d6/pdf} {\bibfield  {journal} {\bibinfo  {journal} {Plasma Phys. Control. Fusion}\ }\textbf {\bibinfo {volume} {62}},\ \bibinfo {pages} {105014} (\bibinfo {year} {2020})}\BibitemShut {NoStop}%
\bibitem [{\citenamefont {Labit}\ \emph {et~al.}(2016)\citenamefont {Labit}, \citenamefont {Nespoli}, \citenamefont {Horacek}, \citenamefont {Tsui}, \citenamefont {Boedo}, \citenamefont {Theiler}, \citenamefont {Furno}, \citenamefont {Halpern}, \citenamefont {Ricci}, \citenamefont {Pitts} \emph {et~al.}}]{labit2016physics}%
  \BibitemOpen
  \bibfield  {author} {\bibinfo {author} {\bibfnamefont {B.}~\bibnamefont {Labit}}, \bibinfo {author} {\bibfnamefont {F.}~\bibnamefont {Nespoli}}, \bibinfo {author} {\bibfnamefont {J.}~\bibnamefont {Horacek}}, \bibinfo {author} {\bibfnamefont {C.}~\bibnamefont {Tsui}}, \bibinfo {author} {\bibfnamefont {J.}~\bibnamefont {Boedo}}, \bibinfo {author} {\bibfnamefont {C.}~\bibnamefont {Theiler}}, \bibinfo {author} {\bibfnamefont {I.}~\bibnamefont {Furno}}, \bibinfo {author} {\bibfnamefont {F.}~\bibnamefont {Halpern}}, \bibinfo {author} {\bibfnamefont {P.}~\bibnamefont {Ricci}}, \bibinfo {author} {\bibfnamefont {R.}~\bibnamefont {Pitts}}, \emph {et~al.},\ }\bibfield  {title} {\bibinfo {title} {The physics of the heat flux narrow decay length in the {TCV} scrape-off layer: experiments and simulations},\ }in\ \href {https://scipub.euro-fusion.org/wp-content/uploads/eurofusion/WPMST1CP16_15302_submitted.pdf} {\emph {\bibinfo {booktitle} {Proceedings of the 26th IAEA Fusion Energy Conference, Kyoto, Japan}}}\ (\bibinfo
  {year} {2016})\BibitemShut {NoStop}%
\bibitem [{\citenamefont {Dejarnac}\ \emph {et~al.}(2015)\citenamefont {Dejarnac}, \citenamefont {Stangeby}, \citenamefont {Goldston}, \citenamefont {Gauthier}, \citenamefont {Horacek}, \citenamefont {Hron}, \citenamefont {Kocan}, \citenamefont {Komm}, \citenamefont {Panek}, \citenamefont {Pitts} \emph {et~al.}}]{dejarnac2015understanding}%
  \BibitemOpen
  \bibfield  {author} {\bibinfo {author} {\bibfnamefont {R.}~\bibnamefont {Dejarnac}}, \bibinfo {author} {\bibfnamefont {P.}~\bibnamefont {Stangeby}}, \bibinfo {author} {\bibfnamefont {R.~J.}\ \bibnamefont {Goldston}}, \bibinfo {author} {\bibfnamefont {E.}~\bibnamefont {Gauthier}}, \bibinfo {author} {\bibfnamefont {J.}~\bibnamefont {Horacek}}, \bibinfo {author} {\bibfnamefont {M.}~\bibnamefont {Hron}}, \bibinfo {author} {\bibfnamefont {M.}~\bibnamefont {Kocan}}, \bibinfo {author} {\bibfnamefont {M.}~\bibnamefont {Komm}}, \bibinfo {author} {\bibfnamefont {R.}~\bibnamefont {Panek}}, \bibinfo {author} {\bibfnamefont {R.}~\bibnamefont {Pitts}}, \emph {et~al.},\ }\bibfield  {title} {\bibinfo {title} {Understanding narrow {SOL} power flux component in {COMPASS} limiter plasmas by use of {Langmuir} probes},\ }\href {https://doi.org/10.1016/j.jnucmat.2014.12.100} {\bibfield  {journal} {\bibinfo  {journal} {J. Nucl. Mater.}\ }\textbf {\bibinfo {volume} {463}},\ \bibinfo {pages} {381} (\bibinfo {year}
  {2015})}\BibitemShut {NoStop}%
\end{thebibliography}%

\end{document}